**Title**

Kinetic Monte Carlo modelling of Helium Bubble Nucleation onto Oxides in the Fe-Ti-Y-O System

**Authors**

Chris Nellis[1], Céline Hin[1,2]

[1] Department of Mechanical Engineering, Virginia Tech, Blacksburg, VA 24061, USA

[2] Department of Material Science and Engineering, Virginia Tech, Blacksburg, VA 24061, USA

# Abstract

A Kinetic Monte Carlo (KMC) model was created to simulate the insertion of transmutation He atoms into nanostructured ferritic alloys (NFAs) under neutron irradiation. Interstitial He atoms migrate through the NFA until becoming trapped in bubbles of other He atoms and vacancies created from irradiation. The Y-Ti-O nano-oxides in the NFAs were found to be effective in capturing these He atoms and preventing bubbles from forming at the grain boundary and appear to replicate the characteristics (size and number density) observed in other experiments. The bubbles were found to prefer the <111> oxide interface as a nucleation site and the stable bubbles have a He/Vac ratio between 1.3 and 1.8 He/Vac. The influence of He bubbles on the segregation of solutes to the grain boundaries or on the stability of the nano-oxides were negligible.

# I. Introduction

Cladding materials inside a nuclear reactor are expected to handle the insertion of He atoms from the (n,α) nuclear reactions taking place in the reactor. A primary mechanism of the degradation of metals in a nuclear reactor is the formation of large He bubbles at the metal grain boundaries. These bubbles subsequently form voids that lead to a loss in yield strength which shortens the lifetime of the material. This weakness needs to be accounted for since the next generation of nuclear reactors will have more intense irradiation environments and helium insertion rates.

The nanostructured ferritic alloys (NFAs) are a proposed class of material designed to mitigate the degradation caused by transmutation He. The goal being to increase resistance to He void swelling which occurs once a He bubble reaches a certain critical size, it will begin to quickly trap free vacancies and grow into a void responsible embrittling the material. The NFAs defining feature is a high number density of very small oxide nanoparticles (<2nm diameter) typically containing Y-Ti-O atoms embedded in the ferritic matrix [1]. Their primary purpose is to provide an alternative nucleation site for the He bubbles as opposed to other defects like the grain boundary. The result is a high density of small He bubbles within the grain which causes less degradation of the material properties than large bubbles at the grain boundary would. So the oxides in the NFAs improve the irradiation resistance in two ways. 1) Keeping the He bubble away from the grain boundary and 2) minimizing the size of the bubbles to delay the formation of voids. These nano-oxides are highly resistant to dissolution at the high operating temperatures [2] and are resistant to irradiation induced dissolution, so the oxides maintain their He capture properties in the reactor environment.

Other advantages to the nano-oxides are their pinning effect that restricts the movement of dislocations, thus improving the mechanical strength of the material [3]. The smaller grain size also increases the grain boundary surface area which reduces the build-up of vacancies in the material by providing more opportunities for defect sinks to annihilate the vacancies [4]. The oxide particles themselves act of defect sinks which would also reduce the defect population.

Research into the NFA's ability to trap He atoms at the oxides is ongoing. There is some difficulty replicating the insertion of transmutation helium into metals in experimental settings outside an actual nuclear reactor. For physical evaluations of the helium bubble distributions, helium implantation techniques have been developed to mimic the irradiation regimes in reactors by having spallation neutrons induce a (n,α) reactions in a blanket material surrounding the sample, thus releasing He atoms into the sample material with great care made to get the correct He-to-dpa ratios [5]. Sometimes the neutrons are substituted with accelerated ions, though the applicability to neutron irradiation is debatable [6]. Odette uses this technique to conduct a helium implantation study that found MA957 is proficient at preventing bubble nucleation at the grain boundary[7]. Two companion papers by Kurtz [5] and Yamamoto [8] observe the He bubble characteristics in non-ODS steel and the NFA MA957 respectively which confirm the NFAs reduce the size of the bubbles and trap the bubbles at oxide sites. A study by Edmondson for the 14YWT NFA found that roughly 20% of the He bubbles reside in the matrix, 49% reside at the nanoclusters, and 14.4% reside at the grain boundaries [9]. When the bubbles do nucleate in an NFA like the 14YWT alloy, the helium bubbles are smaller in size and volume than a conventional steel [10]. Other experiments observe that the helium bubbles tend to nucleate at the <111> interface of the oxides owing to the <111> interface's high interface energy [11]. This also notes that it is difficult to get imaging down to the 2nm scale so modelling of the shape of the oxide and location of the He bubbles using a computer model would be useful.

The stated difficulty of conducting reactor experiments and a need to for a greater understanding of every factor influencing outcomes has encouraged the use of computer models in nuclear materials research. When used in conjunction with experiments, researchers can better determine the precise mechanisms for material behavior [12].

Kinetic Monte Carlo (KMC) has been extensively used to simulate the precipitation under heat treatment[13, 14] and effects of irradiation on the material microstructure [15]. While usually the models are developed for binary or tertiary alloys with substitutional elements, some models include interstitial elements like the models for the Fe-Nb-C [16], Fe-Ti-O [17] and Fe-Y-O [18] systems. There have been some studies of the KMC with interstitial elements including helium in the pure bcc Fe [19-21]. The current KMC model developed in this study is an extension of the model used to study the precipitation and resistance of Y-Ti-O oxides to dissolution [22, 23].

In this study, a previously developed KMC model for neutron irradiation of NFAs in an Fe-Y-Ti-O system was extended to incorporate the insertion of interstitial helium into a model NFA during the neutron bombardment. The size, location, and He/Vac ratio of the He bubbles in the model NFA were noted and compared to experimental results. A preliminary investigation was conducted to observe the effect of temperature on the characteristics of the He bubbles. Investigations into the effect of the He bubbles on oxide stability and the segregation of solutes were also conducted.

# II. Material/Methodology

## 1. KMC Model

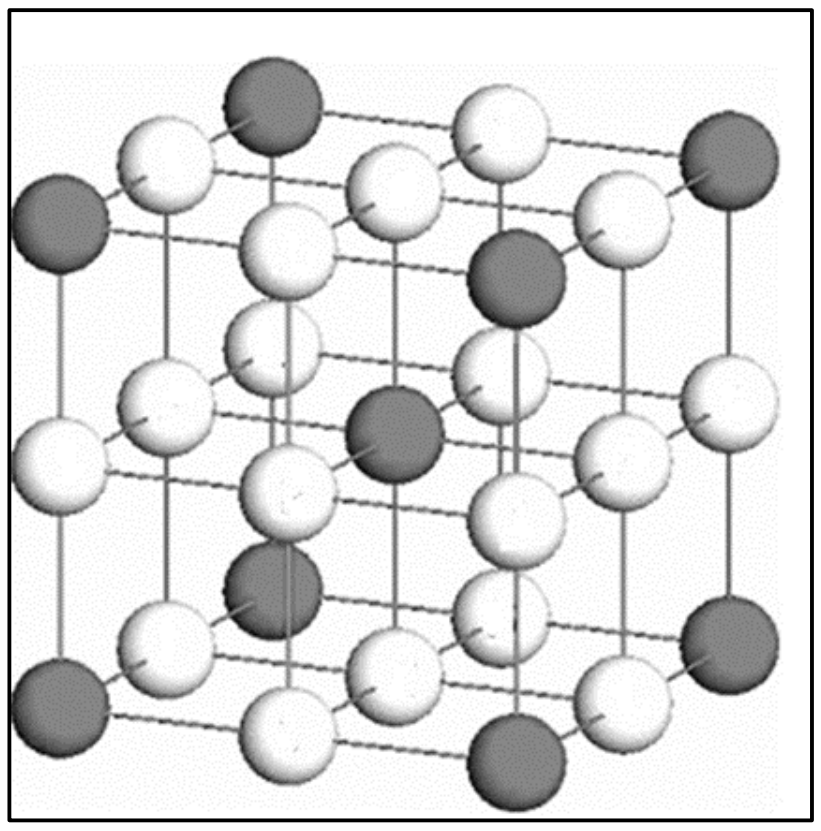

*Figure 1: The lattice of the Fe-Ti-Y-O alloy. The dark circle represents the bcc lattice site for Fe, Y, and Ti atoms and the white circles represent octahedral lattice sites for the O and He atoms.*

The KMC model was extended from a previous model[24] used to study neutron irradiation in NFAs to incorporate the addition of He atoms. A fixed lattice in the bcc system is constructed to represent the bcc cell. The simulation box is populated with atoms at the substitutional sites (Fe, Ti, Y) and at the octahedral sites (O and He) in the bcc system. The simulation box is a rigid lattice system that does not allow for elastic deformation of the lattice from the perfect lattice and thus cannot form a new phase system within the matrix. The small oxides in the NFAs are expected to be small enough not to form their own phase and are instead coherent with the bcc system[25]. The substitutional atoms (Fe, Ti, Y) migrate through exchange mechanisms with the vacancy and interstitial dumbbell point defects. The He and O atoms migrate through an interstitial mechanism to hop between empty interstitial sites. At each Monte Carlo step, an event is performed in the simulation box, either defect production or atomic migration.

$$\Gamma_x = v_x \times \exp(\frac{-E_{mig}}{k_b T}) \quad (1)$$

Each atom migration event is given a jump frequency $\Gamma_x$ that is calculated in Eq. 1 from the migration energy barrier $E_{mig}$ and attempt frequency $v_x$ based around the diffusion properties of each element in bcc Fe. The energy barrier $E_{mig}$ is calculated from the local atomic configuration around the migrating atom.

$$E_{mig} = e^{SP} - \sum_j \varepsilon_{Fej}^3 N_j^3 - \sum_j \varepsilon_{Fej}^4 N_j^4 - \sum_{n=1,2} \varepsilon_{FeO}^n N_O^n - \sum_j \varepsilon_{jV}^3 N_j^3 \quad (2)$$

The local configuration is described by the pair-interaction energies $\varepsilon_{AB}^n$, each pair-interaction representing the interaction between atoms A and B in the n nearest neighbor distance. The model is capable of including up to the 4th nearest neighbor distance in the simple cubic simple (2nd nearest neighbor bcc system). The saddle point energy $e^{SP}$ adjusted to ensure the $E_{mig}$ tied to the recorded migration barrier for the atomic species in pure bcc Fe.

$$\Gamma_{Tot} = \sum \Gamma_x \quad (3)$$

At every Monte Carlo step, all of the event frequencies are compiled into a total frequency $\Gamma_{Tot}$ in Eq 3, and an event is randomly selected using the jump frequencies $\Gamma_x$. Summing each frequency in the event list until the condition $\sum_{i=0}^{n} \Gamma_x < r \times \Gamma_{Tot} < \sum_{i=0}^{n+1} \Gamma_x$, where r is a random number between 0 and 1, is met. Then the event at index *n* is executed. After each step, the jump frequencies are recalculated and the cycle continues until a stop condition is achieved. For this model, the stop condition occurs when the system's total dose, measured in dpa, reaches a user inputted threshold.

Several events are not given a frequency and instead occur as soon as a condition is met. Free vacancies and interstitial dumbbells both recombine and annihilate at grain boundaries instantaneously in the manner described in Soisson[15]. When a vacancy has a He atom as a first-nearest neighbor, that vacancy is not subject to the automatic recombination when the interstitial dumbbell is in the immediate vicinity.

## 2. He Insertion

Unlike the in-situ helium implantation techniques, the He atom is inserted into the simulation box without an energy that could displace atoms. All displacements are due to displacement cascades caused by neutron impacts. The information of the mechanism for the displacement cascades are described in a companion paper [23]. The He implantation rate is tied to the expected He appm/dpa ratio expected from the reactors. Once an appropriate dpa is reached in the simulation, a single He atom is added to a random location in the simulation box. This is a process that occurs automatically and thus is not an event frequency does not need to be established. Literature found the expected He appm/dpa ratio is around 10 He appm/dpa [8], though in this model the He/dpa rate is set to a higher 50 He appm/dpa to see timely formation of He bubbles and to emulate rates from some experimental studies.

The He-Vacancy interactions in the KMC model are expected to be the most influential parameters in the He simulations, since He “bubbles” are really very large He-Vac complexes where the helium atoms take the voided space left by the vacancy clusters. The diffusion properties of the He-Vac complexes are described in a paper by Ortiz [26], although this model treats the migration of He and Vac in clusters as separate entities but are expected to behave the same way. Unlike other KMC models, the clusters are not treated as separate entities with their own migration properties. The association and dissociation events of the individual He atoms with clusters are not given a separate event frequency with a specifically calculated dissociation energy. Rather the local environment is reflected through the calculation of the migration energy of the vacancy using pair-interaction energies.

## 3. Parameterization

### A. Pair-interaction energies

*Table 1: Pair-interaction energies for Fe-Ti-Y-O system (eV) as a function of nearest neighbors*

| | **1 (eV)** | **2 (eV)** | **3 (eV)** | **4 (eV)** |
|---|---|---|---|---|
| Fe-Fe | - | - | -0.611 | -0.611 |
| Fe-Y | - | - | -0.59 | -0.52 |
| Y-Y | - | - | -0.57 | --0.69 |
| Fe-Ti | - | - | -0.65 | -0.53 |
| Ti-Y | - | - | -0.71 | -0.68 |
| Ti-Ti | - | - | -0.69 | -0.70 |
| Fe-Vac | - | - | -0.21 | 0.0 |
| Y-Vac | - | - | -0.35 | 0.0 |
| Ti-Vac | - | - | -0.35 | 0.0 |
| Fe-I | - | - | -0.10 | 0.0 |
| Y-I | - | - | 0.25 | 0.0 |
| Ti-I | - | - | -0.10 | 0.0 |
| Fe-O | 0.00 | 0.00 | - | - |
| Y-O | 0.01 | -0.11 | - | - |
| Ti-O | -0.04 | -0.04 | - | - |
| O-O | 0.10 | -0.116 | 0.10 | -0.116 |
| He-O | -0.34 | | | |
| He-Vac | -2.1 | | | |
| Fe-He | 0.0 | | | |
| Y-He | -0.46 | | | |
| Ti-He | -0.14 | | | |
| He-He | -0.35 | -.42 | | |

Table 1 displays the pair-interaction energies used for the KMC model. The rationale for the non-He interactions are discussed in the companion paper regarding the nucleation of the oxides [22]. There is not a good estimation of the He solubility in bcc Fe so the He-He pair-interaction energies are found using the binding energies of two He atoms in the octahedral sites of pure bcc Fe [27]. The binding energies of the He interstitial with the Ti and Y solutes are described in a paper by Vallinayagam [28] as -0.14 and -0.46 eV respectively. The binding energy for O with He is -0.34 eV. These are used to construct the $\varepsilon^3_{YHe}$, $\varepsilon^3_{TiHe}$, and $\varepsilon^1_{OHe}$ pair-interaction energy with the inherent assumption that all the He interactions can be described by first nearest neighbor interactions. The He-Vacancy interaction energy comes from the binding energy between a single He atom and the vacancy taken from literature[29]. The interaction between the He atom and the interstitial was not included in this model.

A grain boundary was placed in the center and a segregation energy is applied to the migration energy calculation whenever the He atom is seated on the grain boundary. This segregation energy was set to 1.3 eV based on literature data from[30, 31]. Note that unlike the Ti and Y segregation energy, the entropy and enthalpy contributions are not evaluated and the segregation energy is assumed to be constant.

B. Helium Diffusion Properties

For the purposes of the KMC, the helium atoms reside on the octahedral sites like the oxygen atoms. The diffusion of the He atoms are expected to be very quick owing to a very low barrier of migration 0.06 eV. The full characteristics are described in Table 2.

*Table 2: Helium Diffusion characteristics*

| | Pre-exponential | Migration Energy | Source |
|---|---|---|---|
| He interstitial | $2.8 \times 10^{-8}\ m^2/_s$ | 0.064 eV | [32] |

Irradiation simulations from a prior irradiation study are repeated now with the He injection mechanism adding He atoms into the system at regular total dose intervals. The final location and total number of He bubbles are observed and recorded. In addition, the influence of the He bubbles on the stability of the oxides and of segregation of solutes to the grain boundary was studied.

## 4. Simulation Procedure

### A. Interface Study

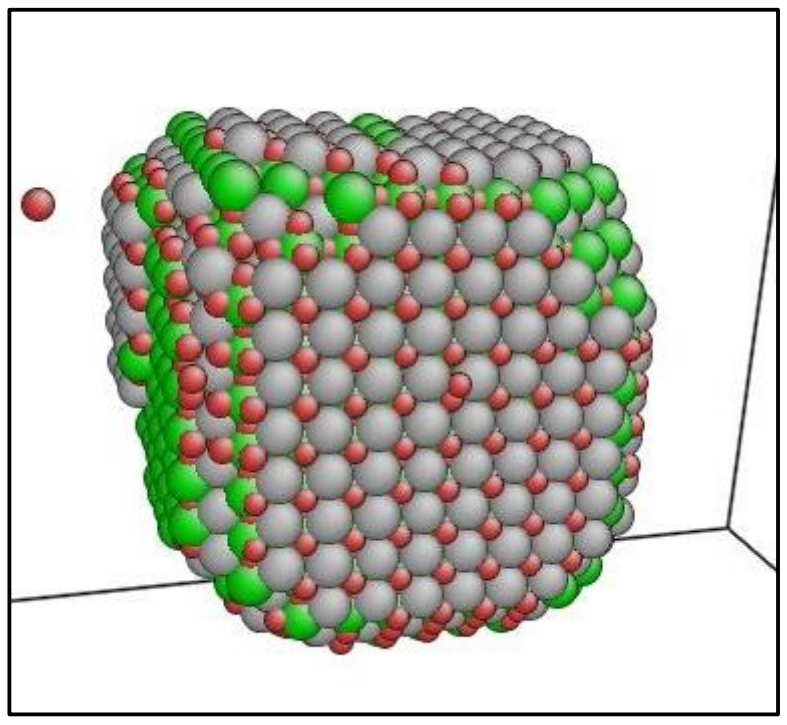

*Figure 2: The equilibrium shape of the Y-Ti-O precipitate at 700K. Green =Ti, Gray=Y, and Red=Oxygen*

In an effort to understand all aspects of He bubble behavior, a focused simulation treatment was designed to study the preferential locations of He bubble nucleation in the KMC model and verify that the behavior matches expectations. Previously conducted simulations in the KMC crafted a large oxide in its thermal equilibrium shape with clearly defined interfaces as seen in Fig 2. The oxides take a cubic shape like observations from Ribis at 1573K [33] with large <100> interfaces and smaller <110> and <111> interfaces. For the interface study, this single large precipitate was placed in a simulation box of pure Fe plus a grain boundary and subject to irradiation conditions listed in Table 3. Since this test part was only interested in the nucleation site, the He appm/dpa rate was set higher than usual to facilitate the timely growth of large He bubbles. After the irradiation, the locations of the bubbles are noted for reference.

*Table 3: Irradiation conditions for the Interface Study.*

| | Size of KMC Box | Temperature | Appm/dpa | Dose Rate | # Simulations |
|---|---|---|---|---|---|
| Interface Study Conditions | 400X100X100 lattice points | 773K | 100 | $10^{-3}\ dpa/s$ | 1 |

B. 14YWT temperature investigation

Another behavior characteristic studied was the influence of irradiation temperature on the size and number density of the He bubbles in the model NFA. For this investigation, the material studied was the 14YWT-1123K using the simulated oxides atomic configuration that was collected in a previous study.

*Table 4: The composition of the 14YWT tested*

| | Y | Ti | O |
|---|---|---|---|
| 14YWT Atomic Concentration | 0.08 at% | 0.27 at% | 0.3 at% |

Table 4 lists the amount of primary components of the nano-oxides (Y,Ti, and O) in the iron matrix of the model NFA to mimic the 14YWT. It's important to note this model representation of the 14YWT does not include elements such as Cr and W. Table 5 lists the characteristics of the oxides after an isothermal heat treatment in the KMC at 1123K.

*Table 5: Oxide characteristics of the 14YWT alloy*

| | Average Radius | Number Density |
|---|---|---|
| 14YWT-1123 K | 1.12 nm | 4.7x10^23 $m^{-3}$ |

Table 6 describes the irradiation conditions simulated in this investigation. The temperatures chosen were 673K, 773K, and 873K. The dose rate chosen was the relatively high $10^{-3}$ dpa/s to due to limitations in the timescale of the simulations. The average size and number density of the bubbles were recorded as well as the locations of the bubbles (on oxides, in the bulk, at grain boundaries). Since the He bubble characteristics between systems would only be comparable with similar amounts of He in the systems, the stopping condition for these simulations has all the simulations achieving the estimated appm He in the simulation box as the reference experiment. In this case, the KMC simulations will stop at 8 dpa when total He concentration will be 400 atomic parts per million (appm).

*Table 6: Irradiation Conditions to observe the formation of He bubbles.*

| | Size of KMC Box | Temperature | Appm/dpa | Dose Rate | # Simulations |
|---|---|---|---|---|---|
| 14YWT | 400X100X100 lattice points | 673K, 773K, 873K | 50 | $10^{-3}\ dpa/s$ | 3 |
| 14YWT | 400X100X100 lattice points | 773K | 50 | $10^{-5}\ dpa/s$ | 3 |
| Pure Fe (No Oxides) | 400X100X100 lattice points | 773 | 50 | $10^{-3}\ dpa/s$ | 3 |

To observe the influence of dose rate on He bubble nucleation, the 14YWT simulations at 773K were repeated at the lower dose rate $10^{-5}$ dpa/s. Additionally, a simulation box of pure Fe is subjected to the 773K $10^{-3}$ dpa/s irradiation environment to observe whether the oxides reduce the size of the He bubbles.  Computational expense prevented the investigation of dose rate at all tested temperatures.

Another aspect of the study was to see the characteristics of the He bubbles. Stable bubbles will maintain a particular ratio of helium atoms and vacancies. While this ratio was not found through physical experiments, there were energetic studies of the He bubbles that estimate the ideal He/Vac ratio [29, 34]. Having the KMC replicate the ideal ratio for the nucleated He bubbles would be another assurance of the model's validity.

C. Microstructure Evolution

The He bubbles could also have an impact on the long-term survivability of the oxides under this irradiation. Therefore, this study also looked the change in oxide size over the course of the neutron irradiation with and without He insertion. For comparison, a 14YWT oxide configuration was subjected the same irradiation environment at 773K but without the He insertion. Any differences in the oxide characteristics will be noted and investigated. Table 7 lists the irradiation conditions. The segregation of the solutes to the grain boundary and the influence of the He bubbles on the segregation was investigated since an area of concern for the embrittlement of NFAs is the degree of radiation-induced solute segregation at the grain boundaries.

*Table 7: Irradiation Conditions to observe the formation of He bubbles.*

| | Size of KMC Box | Temperature | Appm/dpa | Dose Rate | # Simulations |
|---|---|---|---|---|---|
| 14YWT Conditions | 400X100X100 lattice points | 773K | 0.0 | $10^{-3}\ dpa/s$ | 3 |

# III. Results:

## 1. Interface Study

The Y-Ti-O precipitate was placed in a simulation box and subjected to neutron irradiation with the He insertion mechanism activated in the KMC. The simulations continue until several He bubbles were observed on the surface of the oxide. A visual snapshot of the precipitate with He bubbles was created using the software ATOMEYE[35] and inspected for insights into He bubble nucleation behavior.

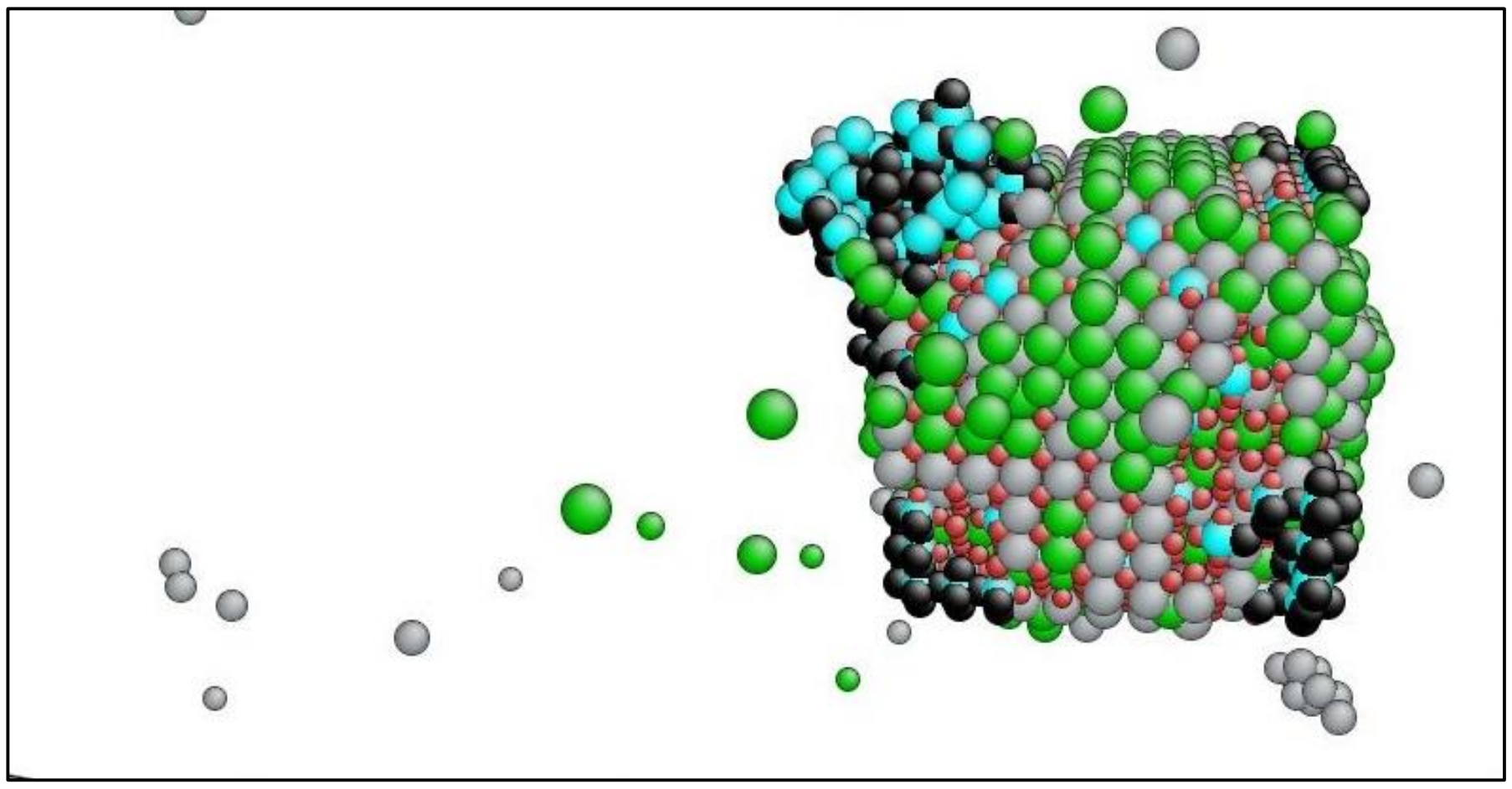

*Figure 3: Helium Bubbles nucleated at the interfaces at the oxide. Green spheres: Ti atoms, Grey spheres: Y atoms, Red spheres: O atoms, Blue spheres: Vacancies, and Black spheres: He atoms*

Fig 3 shows several He bubbles nucleated at the interfaces of the oxide. Most of the He-Vac complexes coated the edges of the oxide with only the largest collection of He and vacancies forming a spherical shape. It appears that the He bubbles prefer to nucleate at the corners of the oxide precipitates at the <111> interface. This result is in line with experimental observations from Stan for He bubbles in 14YWT [11].

## 2. 14YWT Temperature Investigation

After the study of He bubble nucleation behavior on a lone Y-Ti-O oxide, the 14YWT oxide distributions were subjected to neutron irradiation with insertion of transmutation He. The KMC runs simulations at the three temperatures at the same $10^{-3}$ dpa/s dose rate until the system reaches a total irradiation dose of 8 dpa and 400 appm He. The characteristics of the He bubbles were collected and analyzed for any noticeable trends.

### A. Characteristics of He Bubbles

*Table 8: Average size and number density of the He bubbles in the 14YWT after irradiation to 8 dpa*

| | **Average Diameter (nm)** | **Number Density ($m^{-3}$)** |
|---|---|---|
| 673 K $10^{-3}$ $dpa/s$ | 0.68 | $1.8 \times 10^{24}$ |
| 773 K $10^{-3}$ $dpa/s$ | 0.81 | $5.4 \times 10^{23}$ |
| 773 K $10^{-5}$ $dpa/s$ | 0.81 | $5.4 \times 10^{23}$ |
| 873 K $10^{-3}$ $dpa/s$ | 1.46 | $1.42 \times 10^{23}$ |

Table 8 lists the average size and number density of the He bubbles in the 14YWT-like alloy irradiated to 8 dpa. There were roughly 400 appm He atoms in the simulation box of each sample. There was a clear trend in relation to the effect of temperature with the average He bubbles diameter increasing with temperature while the number density declines. At 673K and 773K, the He bubbles nucleate mostly at the interfaces of the nano-oxides and prevent nucleation towards the grain boundary. There is very little difference between dose rates at 773K. Both the $10^{-5}$ and $10^{-3}$ dpa/s case nucleated the same total number of He bubbles across the three simulation runs that averaged to the same size.

*Table 9: KMC results for average size and number density of the He bubbles in the 14YWT and pure Fe after irradiation to 8 dpa at 773K $10^{-3}$ dpa/s.*

| | **Average Diameter (nm)** | **Number Density ($m^{-3}$)** |
|---|---|---|
| 14YWT | 0.81 | $5.4 \times 10^{23}$ |
| Pure Fe (No Oxides) | 1.19 | $1.42 \times 10^{23}$ |

Table 9 shows the difference in He bubble size and density between a simulation box that contains no oxides and one that contains the 14YWT composition. As expected, the 14YWT is shown to precipitate more He bubbles that are smaller in size than those encountered in a system without nano-oxides.

B. 14YWT Experimental Replication

In addition to the simulations attempting to understand the general temperature trends in the 14YWT NFA, we were also interested in the KMC's ability to replicate findings from experimental observations. The findings from the irradiation simulations at 773K were compared against findings from Yamamoto [8].

*Table 10: Comparison of the He bubble characteristics between the KMC and experiment*

| | **Average Diameter (nm)** | **Number Density ($m^{-3}$)** |
|---|---|---|
| From KMC | $0.81$ | $5.4 \times 10^{23}$ |
| From Yamamoto [8] | $0.9 \pm 0.3$ | $3 \times 10^{23}$ |

The distribution of He bubbles in an NFA has been observed experimentally by Yamamoto at 773K in conditions similar to what was conducted. The comparison of these observations and the KMC results are seen in Table 10. There was a close agreement in the bubble size while the KMC has nearly double the number density of He bubbles.

C. Distribution of He Bubbles

To gain a better understanding of the above results, the complete distribution of He bubbles across all the simulations at each temperature were collected and displayed in the following histograms:

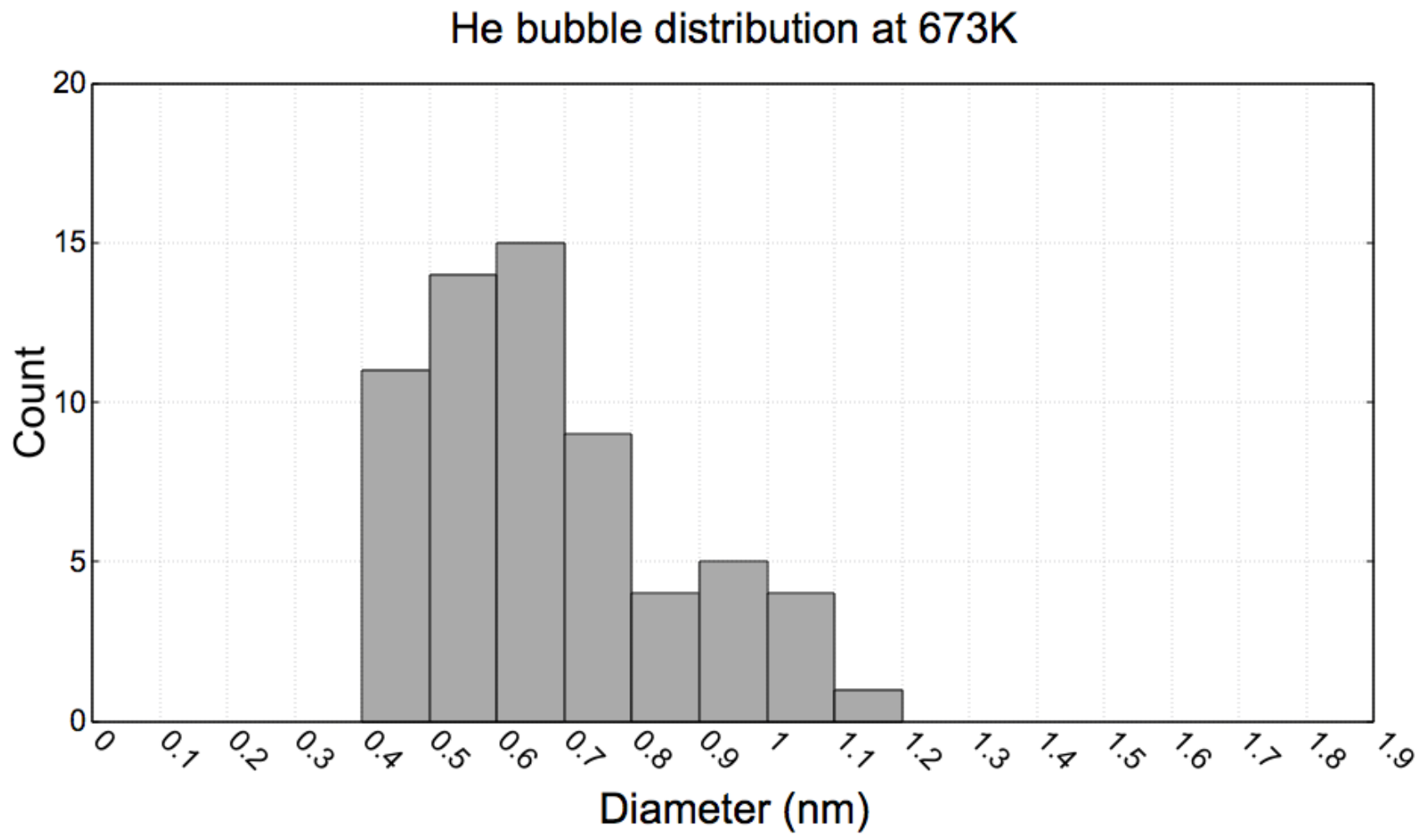

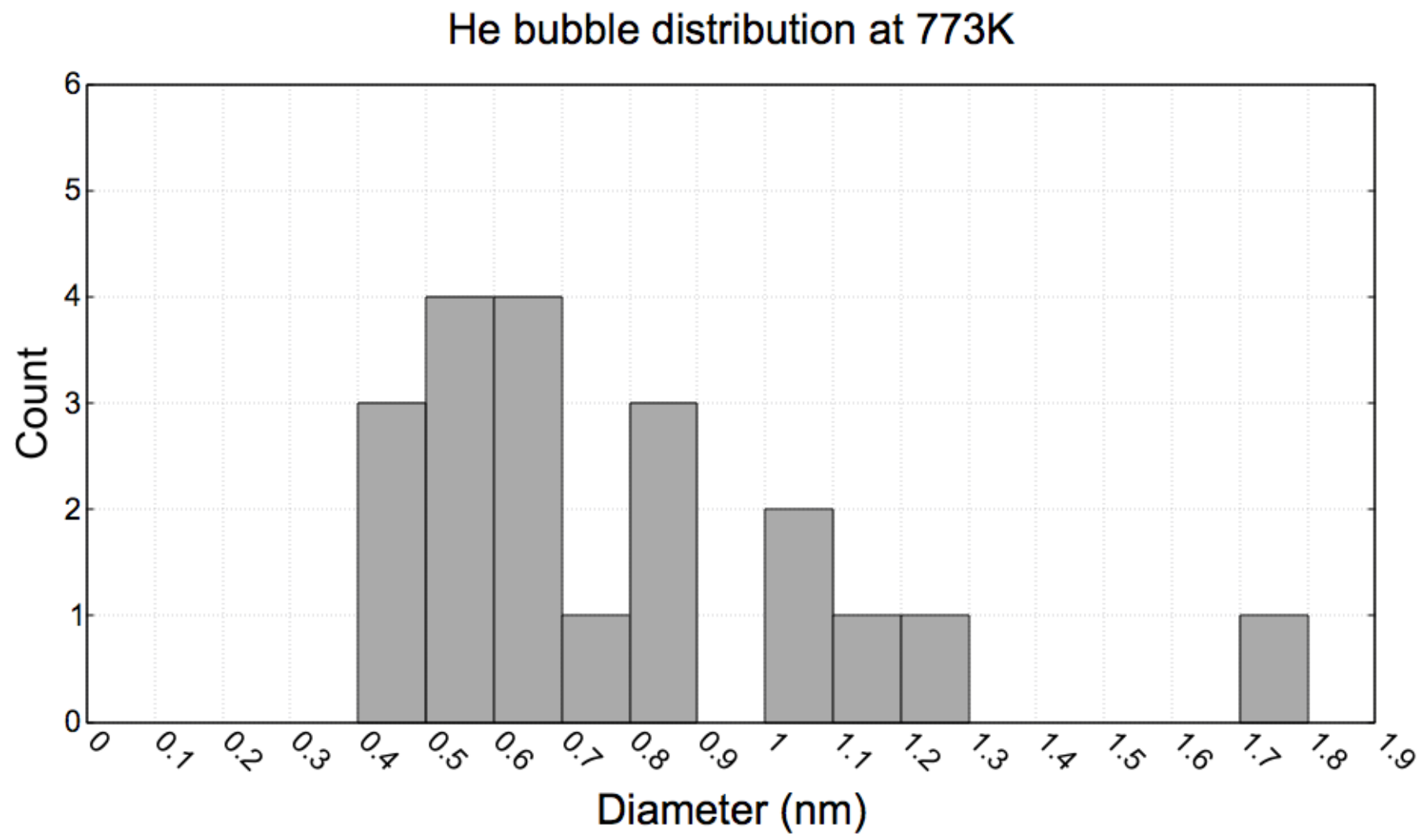

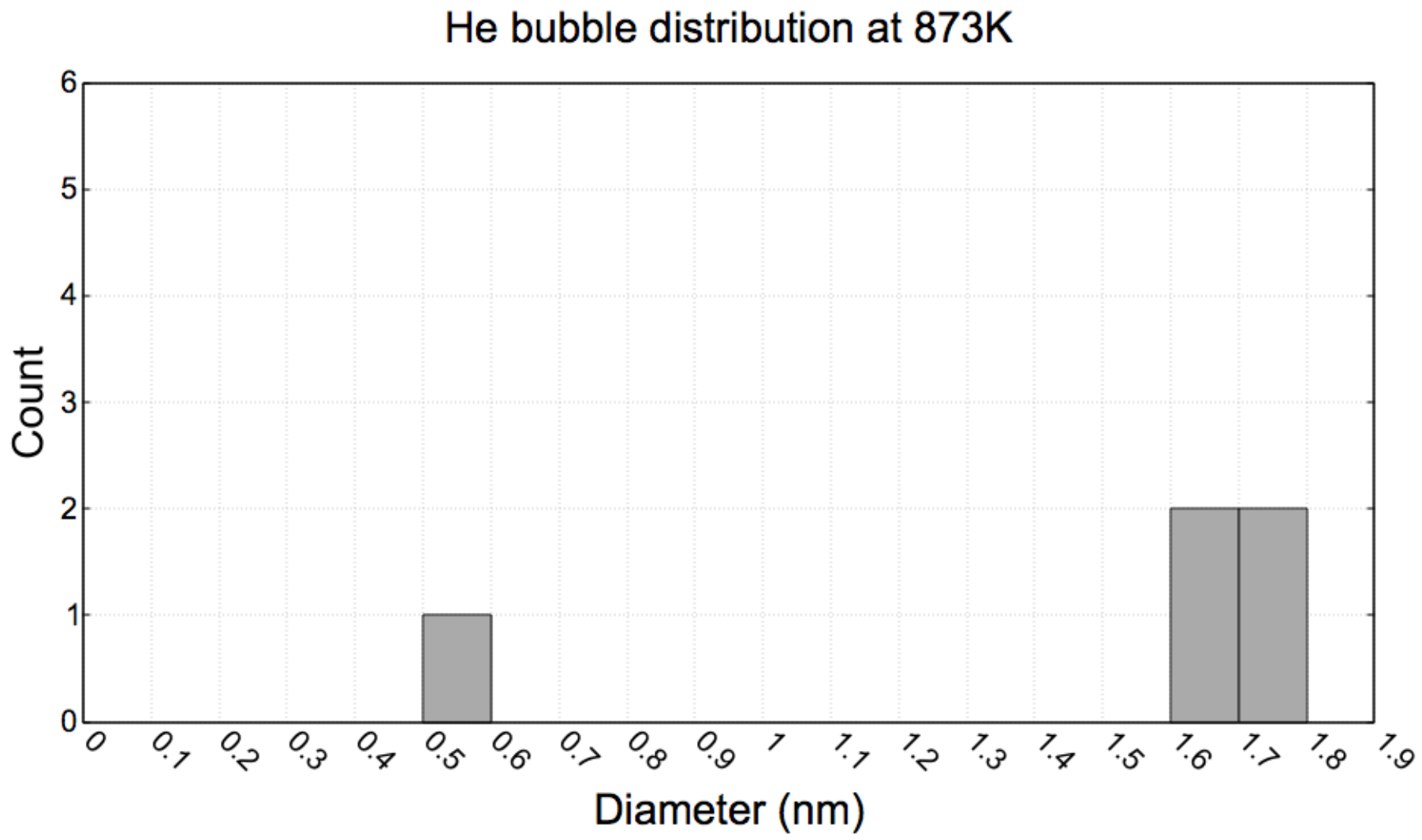

*Figure 4: Helium bubble size distributions after irradiation to 8 dpa at a) 673K b) 773K c) 873K*

Fig 4 shows the size distribution of He bubbles in the simulation box after 8 dpa for the three test temperatures. For the 673K and 773K cases, the highest frequency of bubbles were small bubbles < 1 nm with large >1nm diameter bubbles occurring less frequently. The bubbles in the 673K case were smallest in size with nearly half less than 6 Angstroms in diameter. The bubble size got progressively larger as the simulation temperature was increased with 30% smaller than 0.6 nm for the 773K temperature and only a single bubble under 1 nm for the 873K. The maximum bubble size across all temperatures was 1.8 nm diameter.

D. He/Vac ratio

Another characteristic of the He bubbles observed was the ratio of He atoms to vacancies in the He bubbles for comparison to the ratios expected from literature. The ratio of each bubble was collected and graphed according to their size in order to observe any trends in He/Vac ratio and bubble size.

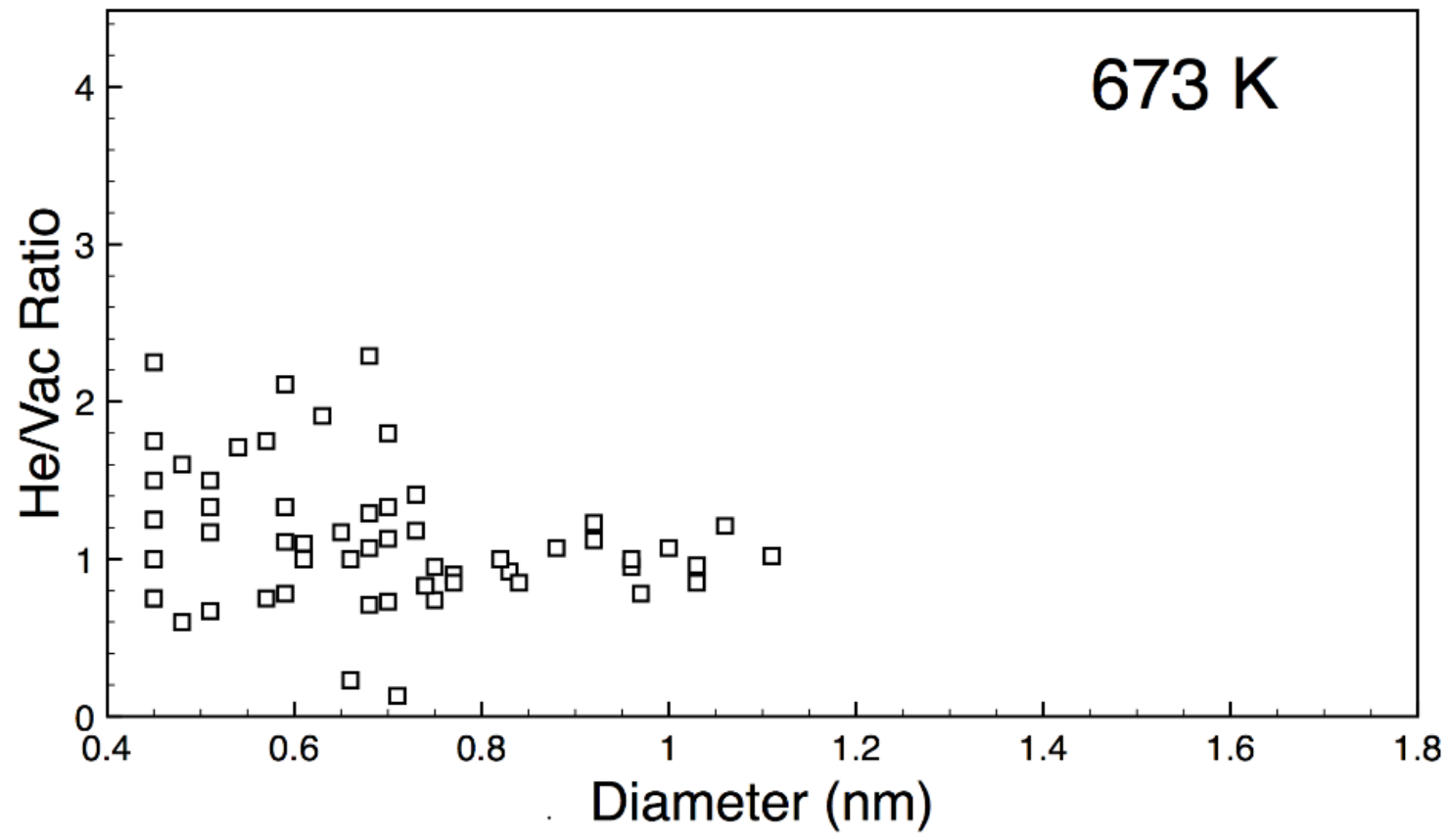

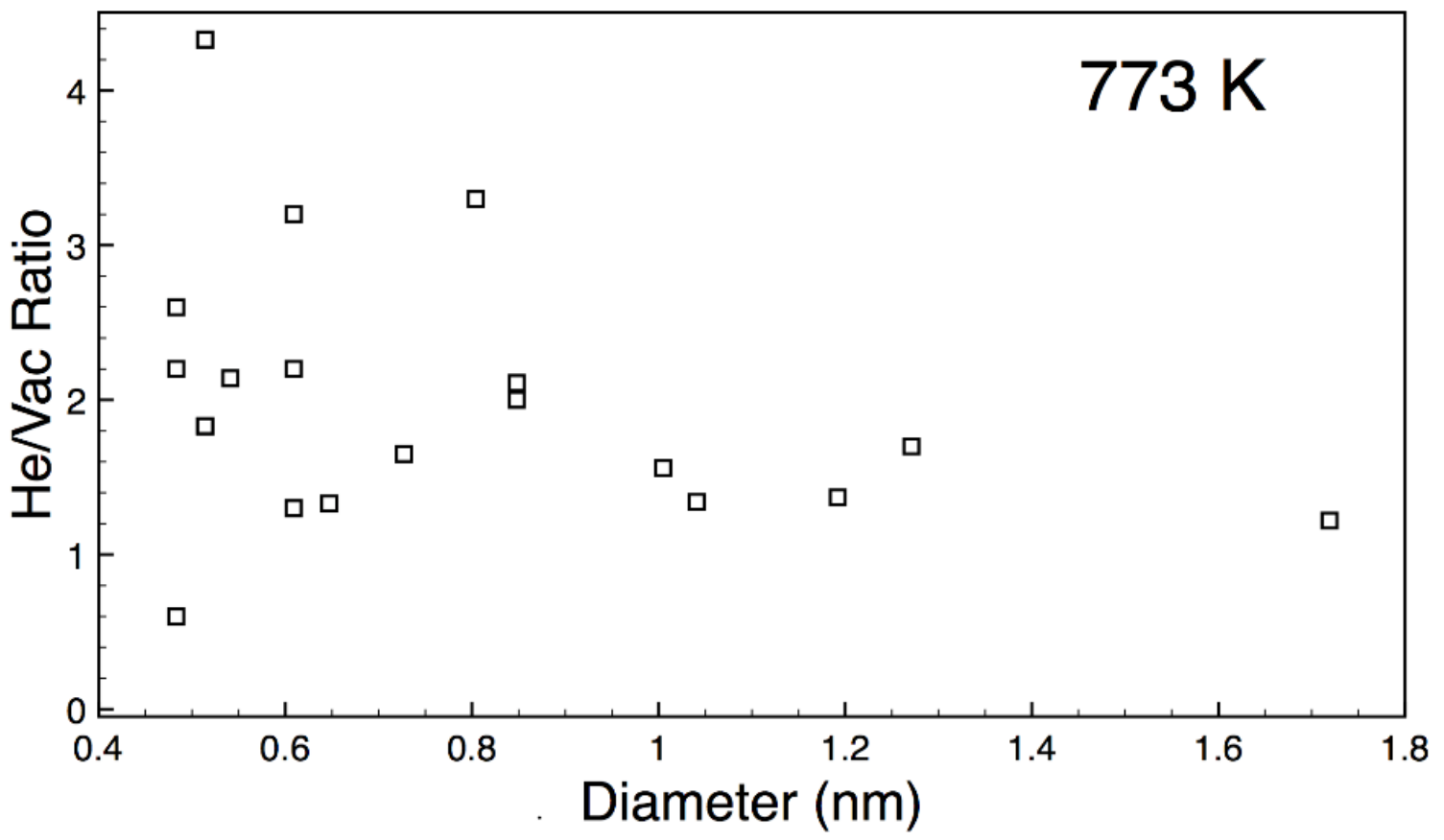

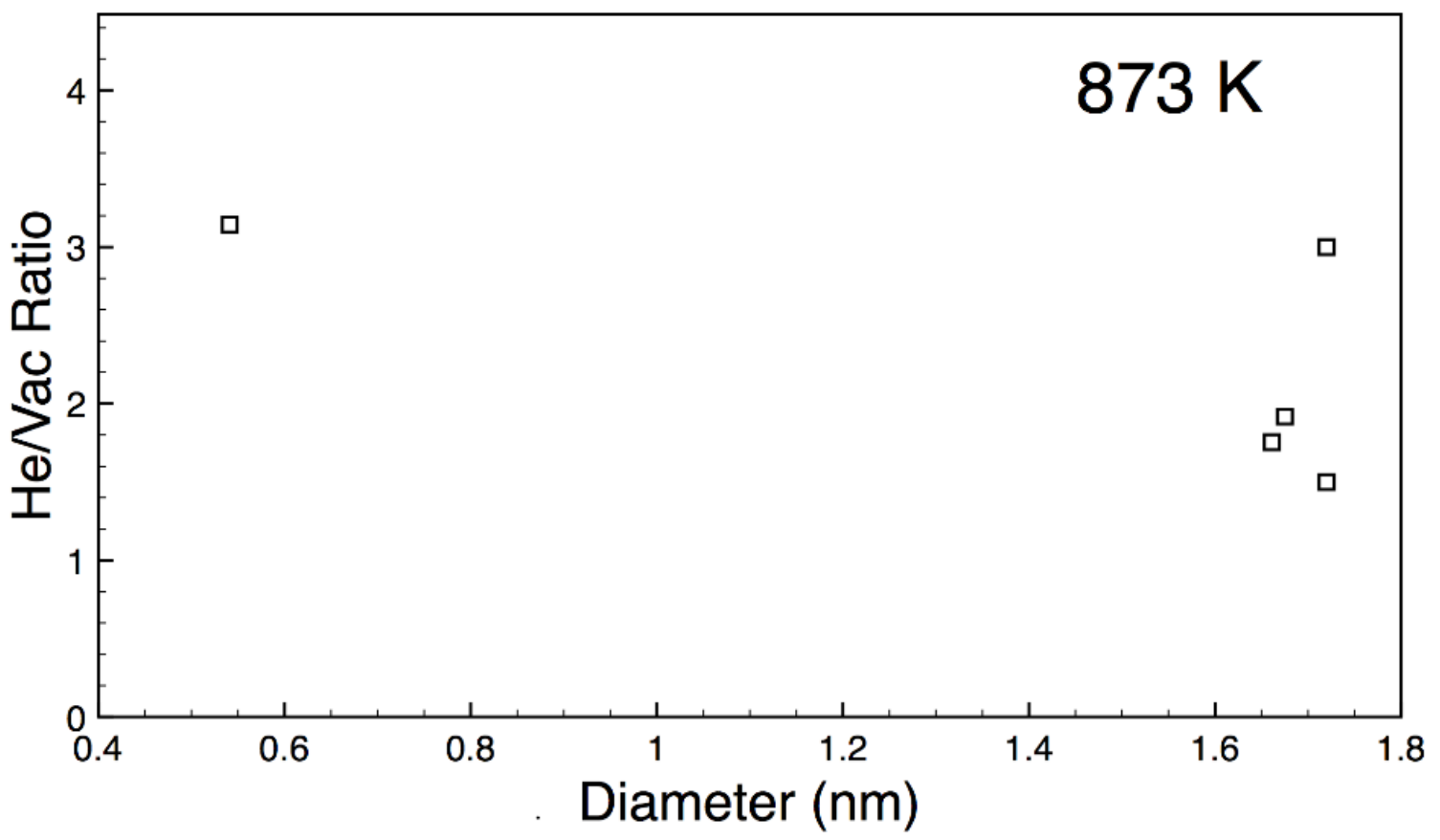

*Figure 5: He/Vac ratio of the He bubbles in the KMC 14YWT at 8 dpa/400 appmHe in irradiation temperatures a) 673K b) 773K c) 873K*

Fig 5 displays the average He/Vac ratio in the He bubbles for each simulation temperature. The majority of the bubbles across all temperatures have a He/Vac ratio that ranges between 1.3 and 1.8. Any bubbles that have ratios outside this range were so different as to be outliers. Many were rather small in comparison to the average bubble size where a change of a single defect/atom will cause a large deviation in the ratio and where the small size could be a factor in the stability. As a He bubble gets larger, the variability decreases with He/Vac ratio appearing to settle into a stable range.

## 3. Evolution of the He Bubble

Insights were gained into the nucleation and growth of the He bubbles. For the first step, the vacancy becomes trapped at the oxide interface with the bcc Fe. Then the He atoms migrating through the matrix binds with this single vacancy. This first initial He bubble is He-rich but as the bubble grows, the ratio of He/Vac reaches a more stable range. There is no formulation of voids during the irradiation simulations.

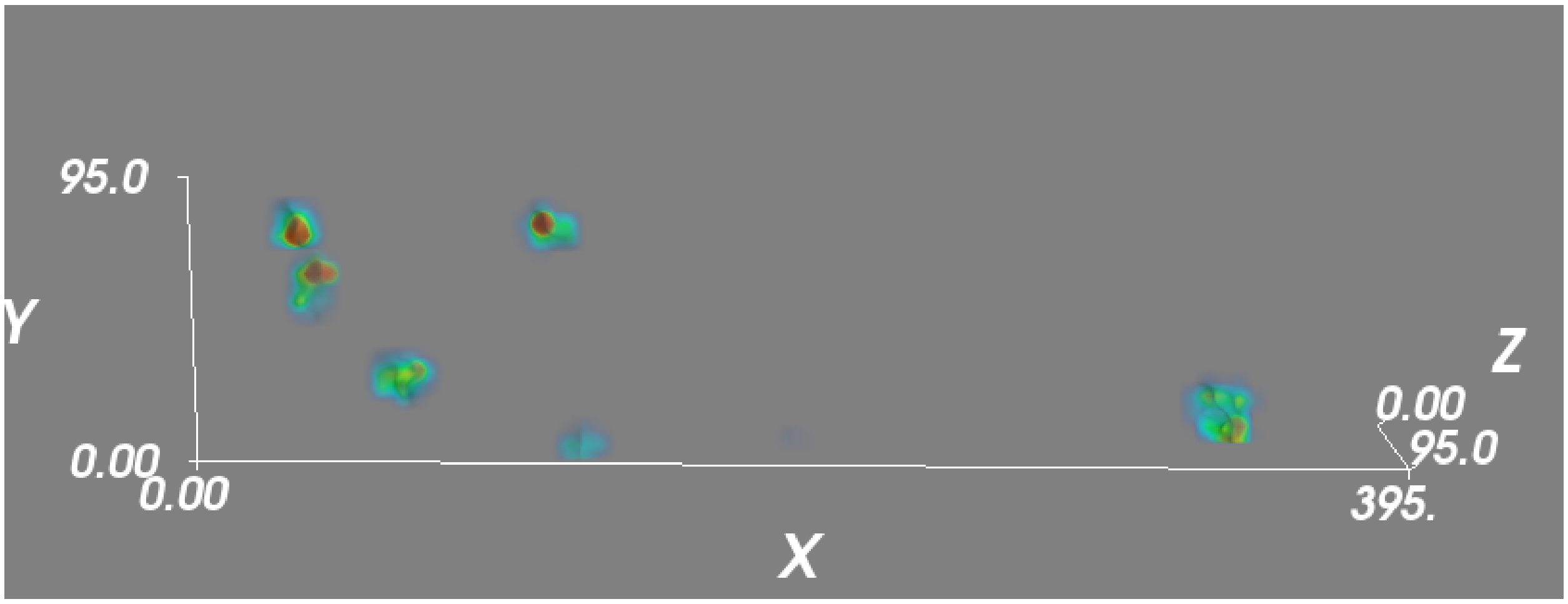

*Figure 6: Heat map of vacancy concentration in 14YWT at 773K irradiation simulations.*

Fig 6 is a heat map of vacancy movement in the simulation box during the irradiation simulations. The areas with the highest average vacancy concentration correspond to the locations of the oxides. With the vacancy spending significantly more time of the interfaces, it is inherently more likely for the He bubbles to form there than in the Fe matrix.

## 4. Effect of He Bubbles on Segregation and Oxide Stability:

A separate irradiation simulation was run alongside the He simulations at 773K with the same irradiation conditions as the 773K case except that no transmutation He was inserted. The concentration profile of the solute atoms in the box was collected and the segregation of the solutes in the grain boundary region of both irradiation cases is noted and compared. The evolution of the oxide size over dpa under the irradiation treatments were also studied in both cases.

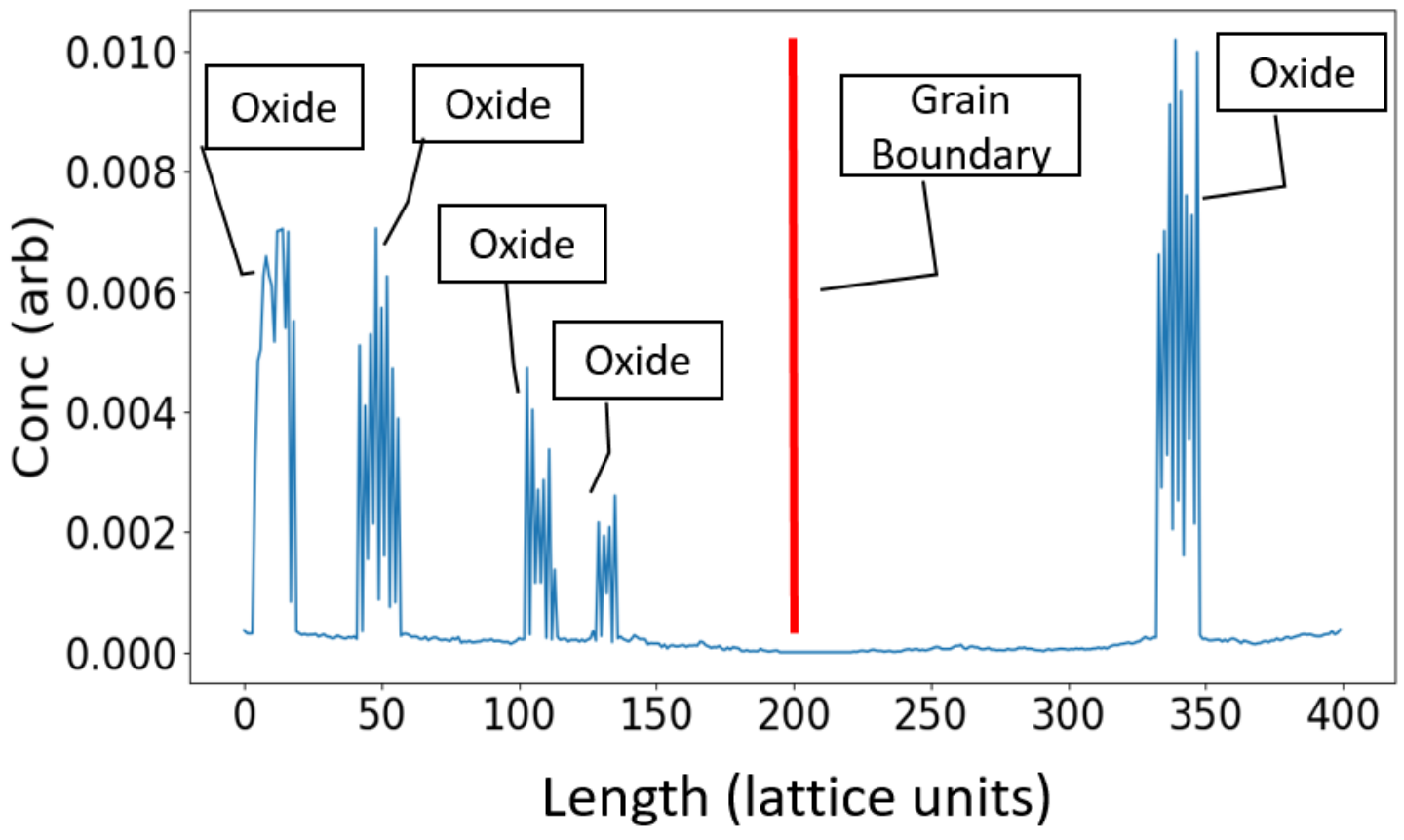

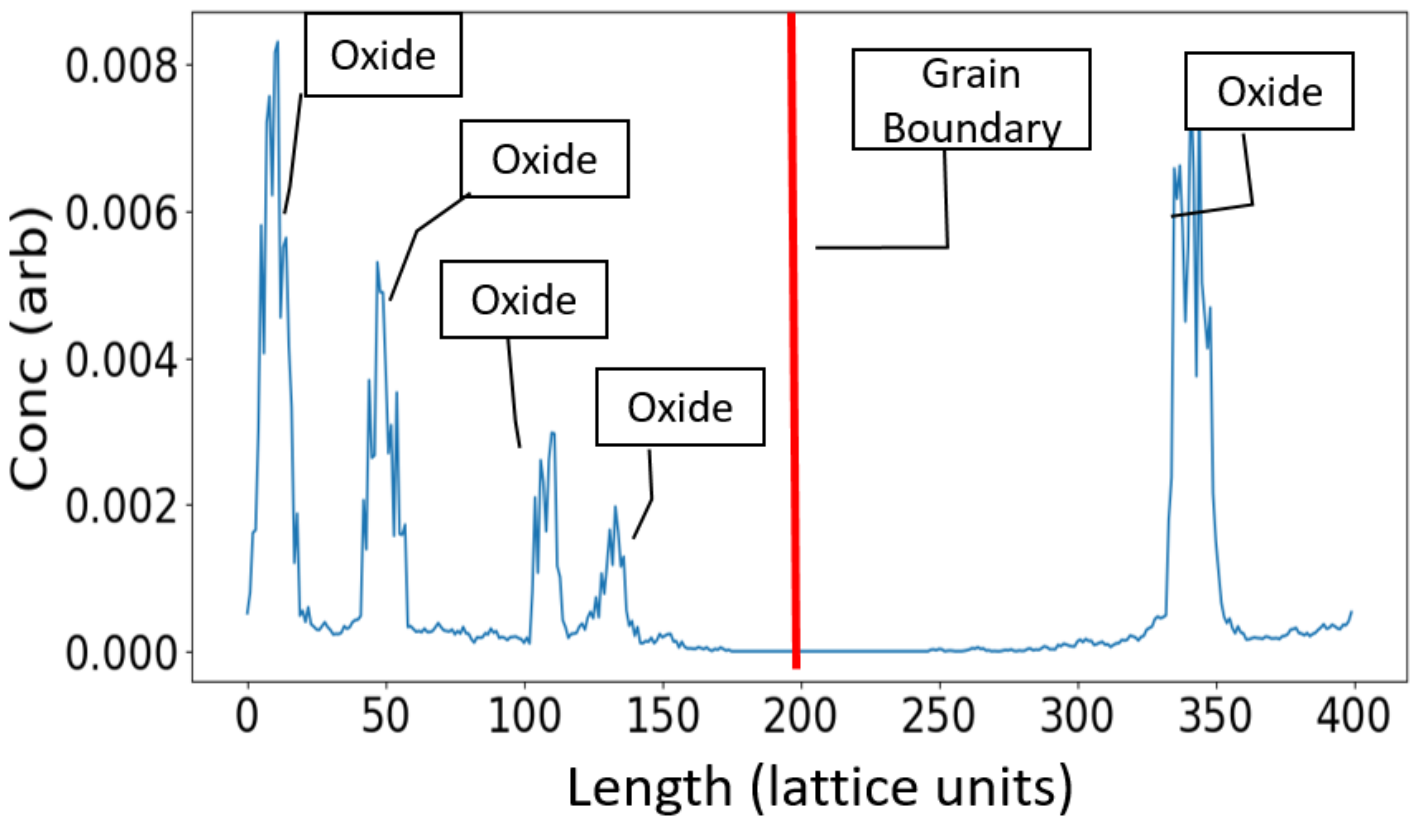

*Figure 7: Yttrium concentration profile of the 14YWT system after irradiation to 8 dpa a) without He insertion b)with He insertion to 400 appm He. The red line represents the grain boundary.*

Fig 7 shows the profile of the Y concentration profiles after irradiation. There does not appear to be any enrichment at the grain boundary sites of the solute atoms of Y and Ti with or without He insertion. Rather there was a depletion of both elements around the grain boundary. Most of the free solutes not associated with an oxide are located in the area immediately around the oxides.

Microstructure evolution

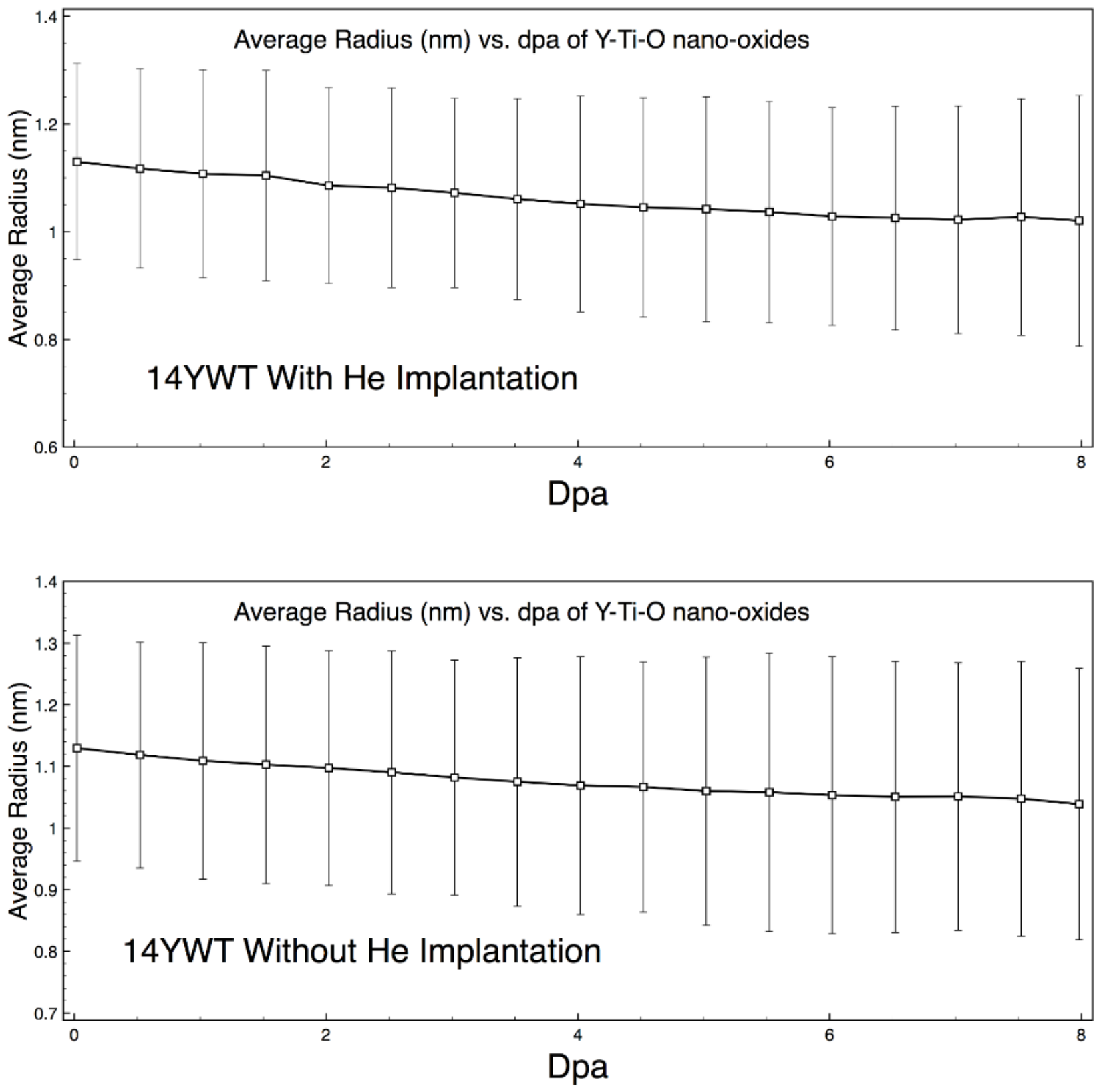

*Figure 8: The microstructure evolution of the 14YWT at 773K and $10^{-3}$dpa/s a) without He insertion and b) with He insertion.*

Fig 8 shows the microstructure evolution of the 14YWT oxides during the irradiation regime up to 8 dpa with and without He insertion. Both cases show a slight decline in the average radius of the oxides during irradiation while there was no decline in the number density and no new precipitates were formed from the ejected solutes in either case. There was no meaningful difference in average radius between the He and non-He cases.

# IV. Discussion

## 1. He Bubble Nucleation Behavior

The oxide particles were shown to act as alternative nucleation sites for He bubbles. First, the interfaces trapped vacancies which served as the first building block of the He bubble The resulting high concentrations of the vacancies around the oxides led to the nucleation and growth He bubbles at the interfaces as it is easier to join an existing bubble than nucleate a new one. There is a lower barrier to nucleation through this path so the He bubble distribution in 14YWT are smaller and more frequent when compared to a pure Fe sample. The interface study in Section III.1 conducted in the beginning sections found the He bubbles preferentially nucleated at the <111> interfaces of the Y-Ti-O oxide. Even the corners without a qualitatively spherical He bubble were found covered in He-Vac complexes. Similar findings were found in experimental results from Stan in an He implantation study, who used larger oxides of the same composition for better imaging [11]. The behavior was also in line with expectations from thermodynamics. The He bubble tends to nucleate at the interface with the highest interface energy and the <111> interface has the largest interface energies for the three primary faces of the Y-Ti-O oxides. The high interface energy of the <111> interface was shown by the small surface area of the <111> interface in the shape of the oxide precipitates, since the oxides reduce their energy by minimizing the surface area of their high energy interfaces.

## 2. Influence of Temperature

The He bubbles observed at 773K were only ~30% the number observed in the 673K simulations while being twice the diameter. A similar difference was seen when comparing the 873K and 773K simulations with the 873K having half the bubble density with 50% larger size by diameter. The observed relation of high number of small He bubbles at low temperatures and small number of larger He bubbles at high temperatures given the same dpa and appm He in the material was backed by literature. In investigations of He bubble characteristics of the non-NFA non-ODS steel Eurofer 97 find substantial differences in average diameter and number density between 673K and 773K [5]. The effectiveness of the nano-oxides in trapping He has not experimentally observed at temperatures beyond 773K [36]. The clear trend of larger bubbles at high temperatures would push the limits of the KMC at those conditions.

When comparing the He bubble characteristics to those found by Yamamoto, the KMC found good agreement with the bubble size though there was a discrepancy with the number density of He bubbles though both were in the same order of magnitude. The discrepancy could simply be due to the limited statistics collected across the 3 simulation boxes. Yamaoto had more bubbles to gather statistics from so more simulations may be needed. The He bubbles of this size were also difficult to image at this size and that difficulty may result in an undercount.

Other areas to consider was the idealized representation of the material system in the KMC model. The NFAs have a high amount of defects and dislocations within the grain that were not represented in the simulation box and other solutes like W and Cr were not included as well. Kurtz finds that the preferred nucleation sites of the He bubbles in a non-NFA ferritic steel at 673K and 773K were at the dislocations that were still plentiful in NFAs alongside the nano-oxides [5]. This could be the source of discrepancies in between results given the absence of dislocation in the simulation box.

The He/Vac ratio of the He bubbles investigated in the KMC results were reasonable with the expectations from literature. The He bubbles at 773K provided the most well defined bubbles for analysis. Once the bubbles reach a certain size, the He/Vac ratio consolidated within a particular bounds of 1.3-1.8 He/Vac. In investigations about the stability of the He bubbles, Morishita [34] found the most stable bubbles energetically have a 1.8 Ha/Vac ratio and Fu's [29] investigation found a He/Vac ratio of 1.3 were the most stable. So the observed He/Vac ratios seen in the KMC were agreeable. The smallest bubbles with less than 10 vacancies have the greatest variability in He/Vac ratio from the oversized influence from their interfaces due to their large relative surface area/volume ratio.

Segregation

The lack of segregation of the solute elements can be attributed to a number of factors. In the beginning of the simulation, the vast majority of solute atoms were incorporated into the oxides owing to the low solubility of all in bcc Fe at the processing temperature. That the irradiation occurs at lower temperatures means that the solutes were even less favorable to be in solution.

The presence of He bubbles in the system does not appear to ultimately affect segregation of the oxide constituent atoms. Most point defects had no interactions with the bubbles before they were annihilated at the grain boundary. The primary cause of separation of the Y and Ti atoms from the oxides was through the ballistic dissolution mechanism. A free solute atom <1nm from the surface of the oxide would likely remain in the neighborhood as the oxides also act as defect sinks. All of the mechanisms that apply to the grain boundary would also apply to the oxide surfaces. Since the inverse Kirkendall effect would keep the Y solutes close to the oxide. This is evidenced by analyzing the composition change in the oxides where there was relatively little decline in the Y component of the oxide in comparison to the Ti content.

Microstructure Evolution

The large oxides of the 14YWT did not dissolve under neutron irradiation for either the He or non-He cases but did experience a slight decline in overall size as expected. There was no radiation induced precipitation of oxides in the bulk. Like with the segregation of solutes, the He bubbles do not appear to have a substantial effect of the stability of the oxides. The point defects continue to migrate through the system relatively unimpeded by the bubbles. It gives more credence to the theory of ballistic dissolution being the cause of oxide shrinkage rather than the increases population of point defects from irradiation. It was possible that further irradiation to higher damage rates would a more drastic impact of the bubbles.

## 3. Next Steps

The next steps for the KMC would be to study the effect of irradiation dose rate on the characteristics of the bubbles if the temperature is held constant. A similar investigation of the effects of the rate of He insertion would also be valuable. The model could also continue to be modified to allow for a larger box size to investigate any effect that might have on the outcome of the He+neutron irradiation particularly with the 873K case.

# V. Conclusion:

A KMC model was created for the purposes of investigating the formation of He bubbles under neutron irradiation typical in an NFA, the primary goal to asses to effectiveness of the nano-oxides in preventing He from reaching the grain boundary. He bubbles formed in this model nucleated and grew on the Y-Ti-O oxides at smaller sizes and a higher number density than when there are no oxides in the system. This outcome is linked to improved resistance to radiation-induced embrittlement. The preferred nucleation site was at the <111> interfaces which in line with expectations from experiments. The KMC model also replicates findings the bubble size and number density under irradiation treatments in prior studies, showing the bubble size increases with increasing temperature. A limited investigation suggested that dose rate has little effect on the resulting He bubble distribution.